\documentclass[twocolumn,aps,pra]{revtex4-2}
\makeatletter
\newcommand*{\rom}[1]{\expandafter\@slowromancap\romannumeral #1@}
\makeatother

\usepackage{color}
\usepackage{graphicx}
\usepackage{hyperref}
\usepackage{amsmath,amssymb} 
\usepackage{epstopdf}
\usepackage{subcaption}
\usepackage{float}
\usepackage{comment}

\graphicspath{{figs/}}

\def\beq{\begin{equation}}
\def\eeq{\end{equation}}
\def\bea{\begin{eqnarray}}
\def\eea{\end{eqnarray}}

\begin{document}
\pagestyle{plain}

\title{Fermionic Unruh effect revisited: Entanglement and correlation dynamics under symmetric transformations}

\author{Omid Yahyayi Monem}
\affiliation{Department of Physics, University of Mohaghegh Ardabili, P.O. Box 179, Ardabil, Iran}
\author{Hosein Mohammadzadeh}
\email{mohammadzadeh@uma.ac.ir}
\affiliation{Department of Physics, University of Mohaghegh Ardabili, P.O. Box 179, Ardabil, Iran}

\begin{abstract}

A uniformly accelerated observer perceives the Minkowski vacuum as a thermal
bath, so acceleration acts as a noise channel on correlations shared with an
inertial partner. For a Dirac field in the standard treatment, the resulting
map is a one-sided amplitude-damping channel, and entanglement and quantum
discord survive at infinite acceleration. We introduce a symmetric Unruh-like
map in which the vacuum and the one-particle state both exchange amplitude with
the inaccessible Rindler region; for a Bell state it yields a bit-flip channel
whose flip probability is the Fermi-Dirac occupation at the Unruh temperature.
Using negativity, mutual information, and all six pairwise quantum discords,
we find that the entanglement and discord between the inertial observer and the
accessible mode vanish together at infinite acceleration, leaving only one bit
of classical correlation. These results show that the fate of quantum
correlations in noninertial frames depends on the structure of the
Minkowski-Rindler map.


\end{abstract}

\maketitle


\section{Introduction}
Relativistic quantum information (RQI) investigates how quantum resources, such as entanglement, coherence, and discord, are affected by relative motion and curved spacetime~\cite{peres2004quantum,alsing2012observer}. A central phenomenon in this context is the Unruh effect, according to which a uniformly accelerated observer perceives the Minkowski vacuum as a thermal state~\cite{unruh1976notes,crispino2008unruh}. This effect arises from the inequivalence of Minkowski and Rindler mode decompositions. Since an accelerated observer cannot access both Rindler wedges, tracing over the inaccessible region transforms an initially pure state into a mixed state and modifies its quantum correlations~\cite{alsing2006entanglement}. The Unruh effect has previously been studied in various respects, including dense coding and teleportation, among others~\cite{mehri2016superdense, mehri2015quantum,farahmand2017residual}. Moreover, the degradation of quantum correlations depends crucially on the field statistics. For bosonic fields, entanglement vanishes only in the infinite-acceleration limit~\cite{fuentes2005alice}, whereas for Dirac fields, fermionic statistics allow a finite amount of entanglement to survive~\cite{alsing2006entanglement}. Similar residual correlations have also been reported for quantum discord and mutual information~\cite{datta2009quantum}. However, the standard single-mode description relies on specific assumptions that may influence these predictions~\cite{bruschi2010unruh}.

From a quantum-channel perspective, tracing over the inaccessible Rindler wedge induces a completely positive trace-preserving map~\cite{bradler2012quantum}. In the standard fermionic treatment, this map possesses an amplitude-damping structure~\cite{nielsen2001quantum}. In contrast, a genuine thermal channel includes both excitation and de-excitation processes~\cite{srikanth2008squeezed}. This physical observation motivates the question of whether an analogous emission-like contribution should also be considered alongside the absorption-like one when modeling the thermalizing effects of acceleration. Addressing this issue from first principles would require a rigorous reformulation within quantum field theory in an accelerated frame, which lies beyond the scope of the present work. We therefore do not modify or rederive the standard Unruh transformation from first principles. Instead, guided by the physical intuition underlying thermal channels, we investigate whether the survival of residual quantum correlations is an inevitable consequence of fermionic statistics or partly an artifact of the asymmetric structure inherent in the standard Unruh transformation.

To explore this question, we introduce a symmetric Unruh-like map as a minimal phenomenological model of acceleration-induced thermalization that incorporates an emission-like contribution in addition to the conventional excitation-like one. Under this transformation, both Minkowski basis states are mapped onto superpositions that distribute amplitudes between the accessible and inaccessible regions in a balanced manner. When applied to an initial Bell state, this transformation produces a tripartite pure state of Alice, the accessible Rindler mode $R$, and the inaccessible mode $\overline{R}$. Tracing out $\overline{R}$ yields an effective bit-flip channel whose flip probability is governed by the Fermi-Dirac occupation number at the Unruh temperature. This construction allows us to systematically examine how the balance between these competing processes affects the ultimate fate of quantum correlations.

We characterize the resulting correlations using negativity and logarithmic negativity~\cite{vidal2002computable}, quantum mutual information~\cite{groisman2005quantum}, and quantum discord~\cite{ollivier2001quantum,henderson2001classical}. All three bipartite reductions are analyzed and compared with their standard-Unruh counterparts. Our results reveal that the entanglement and quantum discord of the accessible pair vanish simultaneously at infinite acceleration, in marked contrast to the standard transformation, for which a finite amount of entanglement persists. Meanwhile, the mutual information approaches one bit, demonstrating that the surviving correlations are purely classical. Moreover, the mutual informations of the accessible and inaccessible pairs satisfy a conservation-like sum rule, while the pairs involving the inaccessible mode remain separable at all accelerations. Of the six pairwise discords, two vanish identically because the corresponding states are classical-quantum, and the discord of the accessible pair is strictly independent of the measurement direction~\cite{luo2008quantum,ali2010quantum}.

The paper is organized as follows. Section~\ref{Thermal Channels} reviews thermal channels. Section~\ref{Unruh effect} presents the standard fermionic Unruh transformation. Section~\ref{Unruh effect revisited} introduces the symmetric transformation and its effective channel. Section~\ref{EUEARUE} examines entanglement, Section~\ref{QMIUEARUE} analyzes quantum mutual information, and Section~\ref{QDUEARUE} investigates quantum discord in both descriptions. Finally, Section~\ref{Conclusion} summarizes the results, discusses the model's limitations, and outlines future directions.


\section{Thermal Channels}\label{Thermal Channels}
In this section, we review the thermal channel, as it constitutes one of the quantum noise models employed in the subsequent analysis. We derive the corresponding Kraus operators and investigate the effect of the thermal channel on a Bell state.

The thermal channel is a finite-temperature generalization of the amplitude-damping (AD) channel~\cite{nielsen2001quantum}, also known as the generalized amplitude-damping channel~\cite{srikanth2008squeezed,khatri2020principles}. Unlike the amplitude-damping channel, which describes spontaneous emission into a zero-temperature environment, the thermal channel accounts for both spontaneous decay and thermally induced excitation arising from the interaction of a qubit with a finite-temperature reservoir. The emission and excitation probabilities, denoted by $p_{\downarrow}$ and $p_{\uparrow}$, respectively, satisfy at thermal equilibrium
\begin{equation}
\frac{p_{\uparrow}}{p_{\downarrow}}=e^{-\hbar\omega/(k_B T)},
\end{equation}
where $\omega$ is the transition frequency, $k_B$ is the Boltzmann constant, and $T$ is the reservoir temperature.

We now consider the Bell state~\cite{nielsen2001quantum}
\begin{equation}
\left|\Phi^+\right\rangle_{AB}
=
\frac{1}{\sqrt{2}}
\left(
|00\rangle + |11\rangle
\right),
\label{Bellstate}
\end{equation}
with the corresponding density operator
\begin{equation}
\rho_{\Phi^+}
=
\left|\Phi^+\right\rangle_{AB}
\left\langle\Phi^+\right|.
\label{DMB}
\end{equation}
The thermal channel for a single qubit admits the Kraus representation~\cite{srikanth2008squeezed}
\begin{equation}\label{TC}
\mathcal{E}_T(\rho)
=
\sum_{i=0}^{2}
K_i \rho K_i^\dagger,
\end{equation}
where the Kraus operators are given by
\begin{equation}
\begin{aligned}
K_0 &=
\begin{pmatrix}
\sqrt{1-p_{\uparrow}} & 0\\
0 & \sqrt{1-p_{\downarrow}}
\end{pmatrix},
\qquad
K_1 =
\begin{pmatrix}
0 & \sqrt{p_{\downarrow}}\\
0 & 0
\end{pmatrix},
\\[6pt]
K_2 &=
\begin{pmatrix}
0 & 0\\
\sqrt{p_{\uparrow}} & 0
\end{pmatrix}.
\end{aligned}
\label{krausG}
\end{equation}
To examine the effect of the thermal channel on the Bell state, we apply the channel to the second qubit, while the first qubit remains unaffected. Using Eqs.~(\ref{DMB}), (\ref{TC}), and (\ref{krausG}), the resulting density matrix is given by
\begin{equation}
\rho'_{AB}
=
(\mathbb{I}\otimes\mathcal{E}_T)
\left(\rho_{\Phi^+}\right)
=
\sum_{i=0}^{2}
(\mathbb{I}\otimes K_i)
\rho_{\Phi^+}
(\mathbb{I}\otimes K_i)^\dagger.
\end{equation}

The resulting density matrix is:
\begin{equation}
\rho'_{AB}
=
\frac{1}{2}
\begin{pmatrix}
1-p_{\uparrow} & 0 & 0 &
\sqrt{(1-p_{\uparrow})(1-p_{\downarrow})}
\\
0 & p_{\uparrow} & 0 & 0
\\
0 & 0 & p_{\downarrow} & 0
\\
\sqrt{(1-p_{\uparrow})(1-p_{\downarrow})}
& 0 & 0 & 1-p_{\downarrow}
\end{pmatrix}.
\end{equation}
In the high-temperature limit, one has $p_{\uparrow} \approx p_{\downarrow} = p$, and the Kraus operators reduce to
\begin{equation}\label{KrausT}
\begin{aligned}
K_0 &=
\begin{pmatrix}
\sqrt{1-p} & 0\\
0 & \sqrt{1-p}
\end{pmatrix},
\qquad
K_1 =
\begin{pmatrix}
0 & \sqrt{p}\\
0 & 0
\end{pmatrix},
\\
K_2 &=
\begin{pmatrix}
0 & 0\\
\sqrt{p} & 0
\end{pmatrix}.
\end{aligned}
\end{equation}
Using Eqs.~(\ref{DMB}), (\ref{TC}), and (\ref{KrausT}), the density matrix of the system is obtained as
\begin{equation}
\rho'_{AB}
=
\frac12
\begin{pmatrix}
1-p & 0 & 0 & 1-p\\
0 & p & 0 & 0\\
0 & 0 & p & 0\\
1-p & 0 & 0 & 1-p
\end{pmatrix},
\end{equation}
written in the computational basis $\{|00\rangle,|01\rangle,|10\rangle,|11\rangle\}$.


\section{Unruh effect}\label{Unruh effect}

In this section, we investigate the Unruh effect and its impact on quantum entanglement, as well as the behavior of quantum states under noise induced by accelerated motion.
The observer-dependence of the particle concept for uniformly accelerated
observers underlies a broad class of phenomena in relativistic quantum
information~\cite{unruh1976notes,crispino2008unruh}. When a bipartite state shared by two inertial parties is
examined from the point of view of an accelerated observer, the mode basis
adapted to that observer no longer coincides with the inertial one, and part
of the correlations available in the inertial description becomes hidden
behind a causal horizon. This redistribution of entanglement was first
studied for bosonic fields~\cite{fuentes2005alice} and later extended to the fermionic case~\cite{alsing2006entanglement,martin2011redistribution}, and is
the physical setting we adopt here.

We consider two observers, Alice and Rob, sharing a bipartite state of a free
Dirac field. Alice remains inertial, while Rob moves with constant proper
acceleration $a$ along the $z$-direction. Rob's natural coordinates are the
Rindler coordinates $(\tau,\xi)$, related to the Minkowski coordinates
$(t,z)$ by
\begin{equation}
(z,t)=\pm\left(\frac{e^{a\xi}}{a}\cosh(a\tau),\ \frac{e^{a\xi}}{a}\sinh(a\tau)\right).
\end{equation}
The two signs define two disjoint coordinate patches Rindler regions I and
II separated by an event horizon: no signal originating in one wedge can
reach the other~\cite{crispino2008unruh}. Because Rob, being confined to one wedge, has no causal
access to the modes of the opposite wedge, the Minkowski vacuum will appear
to him as a mixed state populated by Rindler excitations~\cite{unruh1976notes}. For a spin-$1/2$ field this is made precise starting from the free Dirac
equation,
\begin{equation}
(i\gamma^\mu\partial_\mu-m)\psi=0,
\end{equation}
whose solutions admit a mode expansion in the Minkowski basis natural to
Alice,
\begin{equation}
\begin{aligned}
\psi = \int\left(a_k\,\psi_k^{+}+b_k^{\dagger}\,\psi_k^{-}\right)dk,
\\
\{a_i,a_j^{\dagger}\} = \{b_i,b_j^{\dagger}\}=\delta_{ij}.
\end{aligned}
\end{equation}
or, equivalently, in the Rindler basis natural to Rob, where positive- and
negative-frequency modes must be defined separately in each causally
disconnected wedge,
\begin{equation}
\psi=\int\sum_{\sigma=I,II}\left(c_k^{\sigma}\,\psi_k^{\sigma+}+d_k^{\sigma\dagger}\,\psi_k^{\sigma-}\right)dk .
\end{equation}

Since the two bases correspond to inequivalent notions of positive frequency,
the Minkowski and Rindler operators are related by a Bogoliubov
transformation that mixes particle- and antiparticle-type excitations. For
the Dirac field this transformation reads:
\begin{equation}
\begin{aligned}
a_k &= \cos r\, c_k^{I}-\sin r\, d_{-k}^{II\dagger},\\
b_k &= \cos r\, d_k^{I}+\sin r\, c_{-k}^{II\dagger}.
\end{aligned}
\end{equation}
where the acceleration parameter $r$ is defined through
\begin{equation}
\begin{aligned}
\cos r &= \frac{1}{\sqrt{1+e^{-2\pi\omega c/a}}},\\
\sin r &= \frac{e^{-\pi\omega c/a}}{\sqrt{1+e^{-2\pi\omega c/a}}}.
\end{aligned}
\end{equation}
Here, $c$ is the speed of light and $a$ is Rob's proper acceleration. The frequency $\omega$ is defined in terms of the particle mass $m$ and the momentum vector $\mathbf{k}$ as
\begin{equation}
\omega = \sqrt{|\mathbf{k}|^2+m^2}.
\end{equation}
This relation shows that, from Rob's perspective, a Minkowski particle mode
is unavoidably a superposition of a region-I particle mode and a region-II
antiparticlemode the origin of the acceleration-induced
entanglement between the two wedges. The parameter $r$ interpolates between
the inertial limit $r\to0$ and the infinite-acceleration limit $r\to\pi/4$,
at which particle antiparticle mixing becomes maximal.

Restricting the analysis to the single Rindler mode pair Bogoliubov-related
to a given Minkowski frequency the single-mode approximation commonly
adopted in this literature~\cite{alsing2006entanglement,bruschi2010unruh} the Minkowski vacuum and one-particle states
can be written directly in the Rindler I-II product basis:
\begin{equation}
\begin{aligned}
|0\rangle_M &= \cos r\,|0\rangle_I|0\rangle_{II}
+\sin r\,|1\rangle_I|1\rangle_{II},\\
|1\rangle_M &= |1\rangle_I|0\rangle_{II}.
\end{aligned}
\end{equation}

Since Rob has no access to the degrees of freedom of region II, the state he
perceives must be obtained by tracing them out the mechanism responsible
for the entanglement degradation (and, for fermions, its partial survival at
infinite acceleration)~\cite{alsing2006entanglement,martin2011redistribution}. Tracing out region II defines a channel acting on Rob's qubit, which admits
the Kraus representation
\begin{equation}
\mathcal{E}_r(\rho)=\sum_{i=0}^{1}K_i\,\rho\,K_i^{\dagger},
\end{equation}
The Kraus operators are given by the following expressions:
\begin{equation}
K_0=\begin{pmatrix}\cos r & 0\\ 0 & 1\end{pmatrix},
\qquad
K_1=\begin{pmatrix}0 & 0\\ \sin r & 0\end{pmatrix},
\end{equation}
and satisfy $K_0^\dagger K_0+K_1^\dagger K_1=\mathbb{I}$. Applying this
channel to the second qubit of the Bell state $\rho_{\Phi^+}$ in
Eq.~(\ref{DMB}) gives
\begin{equation}
\rho'_{AB}=(\mathbb{I}\otimes\mathcal{E}_r)(\rho_{\Phi^+})
=\sum_{k=0}^{1}(\mathbb{I}\otimes K_k)\,\rho_{\Phi^+}\,(\mathbb{I}\otimes K_k)^\dagger
\end{equation}
in the basis $\{|00\rangle,|01\rangle,|10\rangle,|11\rangle\}$.

Here, states in region~I are accessible to the accelerated observer, whereas those in region~II are causally inaccessible. This bipartite structure reflects the observer-dependent notion of vacuum and particle content in relativistic quantum field theory. The effect of the Unruh effect on various measures of quantum correlations including entanglement, quantum mutual information, and quantum discord has been investigated in previous studies~\cite{alsing2006entanglement,mehri2011pseudo}. However, the standard Unruh transformation leads to some counterintuitive behaviors. For instance, in the infinite-acceleration limit, the entanglement between the accelerated and the inertial observer does not vanish completely but instead retains a finite value~\cite{alsing2006entanglement,mehri2011pseudo}. In contrast, for a bosonic mode the entanglement vanishes in the limit of infinite acceleration~\cite{fuentes2005alice}. This behavior raises questions regarding the description of quantum correlations in non-inertial frames.

In this work, we introduce a revised description of the Unruh effect by modifying the transformation of the Minkowski vacuum and one-particle states. The purpose of this modification is to re-examine the effect of acceleration on quantum correlations and to address the counterintuitive behavior associated with the standard transformation in particular, the persistence of entanglement in the infinite-acceleration limit. In the following section, we investigate this revised Unruh effect and analyze its consequences for quantum correlations.

\section{Unruh effect revisited}\label{Unruh effect revisited}

In this section, we formulate the symmetric Unruh-like map for the accelerated qubit, serving as the core of our phenomenological framework. Unlike the standard Unruh transformation, in which the vacuum and excited states transform asymmetrically, we introduce a symmetric mapping that incorporates balanced contributions from both excitation-like and emission-like processes. To investigate the impact of this structure on quantum correlations, we revisit an initial Bell state shared between an inertial observer, Alice, and a uniformly accelerated observer, Rob. We replace Rob's Minkowski basis states with the following symmetric transformation:

\begin{align}
|0\rangle_M &\rightarrow 
\cos r\, |0\rangle_R|0\rangle_{\overline{R}}
+
\sin r\, |1\rangle_R|1\rangle_{\overline{R}},
\nonumber\\
|1\rangle_M &\rightarrow 
\cos r\, |1\rangle_R|0\rangle_{\overline{R}}
+
\sin r\, |0\rangle_R|1\rangle_{\overline{R}}.
\end{align}

Under this transformation, the Minkowski vacuum state is mapped to a superposition of particle and vacuum components distributed between Rindler wedges I and II, similarly to the standard scenario. The key departure lies in the transformation of the single-particle state, which in the present formulation assigns symmetric amplitudes to both the accessible and inaccessible regions, thereby reflecting both excitation and de-excitation channels on an equal footing. 

Applying this transformation to Rob's qubit in the initial Bell state introduced in Eq.~(\ref{Bellstate}), the resulting tripartite pure state in the joint Hilbert space $\mathcal{H}_A\otimes \mathcal{H}_R\otimes \mathcal{H}_{\overline{R}}$ is obtained as

\begin{equation}
\begin{aligned}
|\Phi^{+}\rangle_{A,R,\overline{R}}^{(\mathrm{sym})}
&=
\frac{1}{\sqrt{2}}
\Big(
|0\rangle_A
\left(\cos r\, |0\rangle_R |0\rangle_{\overline{R}}
+
\sin r\, |1\rangle_R |1\rangle_{\overline{R}}\right)\\
&\qquad\quad+
|1\rangle_A
\left(\cos r\, |1\rangle_R |0\rangle_{\overline{R}}
+
\sin r\, |0\rangle_R |1\rangle_{\overline{R}}\right)
\Big).
\end{aligned}
\end{equation}

Since the subsystem $\overline{R}$ corresponds to the inaccessible degrees of freedom residing in the opposite Rindler wedge, the physical state accessible to Alice and Rob is obtained by tracing out this unobservable sector:

\begin{equation}
\rho_{A,R}^{(\mathrm{sym})}
=
\mathrm{Tr}_{\overline{R}}
\left(
|\Phi^{+}\rangle_{A,R,\overline{R}}^{(\mathrm{sym})}
\langle\Phi^{+}|_{A,R,\overline{R}}^{(\mathrm{sym})}
\right).
\end{equation}

Carrying out this partial trace explicitly separates the contributions associated with the two Rindler sectors, yielding the reduced density matrix

\begin{equation}\label{rhoarsy}
\rho_{A,R}^{(\mathrm{sym})}
=
\frac{1}{2}
\begin{pmatrix}
\cos^{2} r & 0 & 0 & \cos^{2} r \\
0 & \sin^{2} r & \sin^{2} r & 0 \\
0 & \sin^{2} r & \sin^{2} r & 0 \\
\cos^{2} r & 0 & 0 & \cos^{2} r
\end{pmatrix}.
\end{equation}

To establish a direct connection with the thermal parameters of the Unruh effect, we define the excitation probability as

\begin{equation}
p(r)=\sin^{2}r
= \frac{1}{e^{2\pi\omega/a}+1},
\end{equation}

while the corresponding de-excitation probability is given by

\begin{equation}
1-p(r)=\cos^{2}r
=
\frac{e^{2\pi\omega/a}}
{e^{2\pi\omega/a}+1}.
\end{equation}

These probabilities reproduce precisely the Fermi-Dirac occupation numbers dictated by the Unruh temperature. Consequently, the reduced density matrix can be expressed in terms of the thermal parameter $p(r)$ as

\begin{equation}
\rho_{A,R}^{(\mathrm{sym})}
=
\frac{1}{2}
\begin{pmatrix}
1-p(r) & 0 & 0 & 1-p(r) \\
0 & p(r) & p(r) & 0 \\
0 & p(r) & p(r) & 0 \\
1-p(r) & 0 & 0 & 1-p(r)
\end{pmatrix}.
\end{equation}

This representation highlights the fundamental structural difference between our symmetric Unruh channel and the conventional one. In the standard fermionic treatment, the resulting noise channel is asymmetric and predominantly affects a single transition sector, resembling an amplitude-damping channel. In contrast, the symmetric transformation incorporates balanced contributions from both excitation and de-excitation processes. 

It is worth noting, however, that unlike a standard thermal relaxation channel, the resulting density matrix also retains non-zero coherence terms between the states $|01\rangle$ and $|10\rangle$. These coherences originate directly from the coherent superpositions in the symmetric mapping, indicating that the effective dynamics cannot be reduced to a purely incoherent thermal process. Instead, the net transformation governing the reduced state behaves precisely as a bit-flip channel with an acceleration-dependent flip probability. 

The corresponding Kraus operators for this effective channel can be written as
\begin{equation}
\begin{aligned}
K_0(r) &= \sqrt{1-p(r)}\,I
=
\begin{pmatrix}
\sqrt{1-p(r)} & 0\\
0 & \sqrt{1-p(r)}
\end{pmatrix},\\
K_1(r) &= \sqrt{p(r)}\,X
=
\begin{pmatrix}
0 & \sqrt{p(r)}\\
\sqrt{p(r)} & 0
\end{pmatrix},
\end{aligned}
\end{equation}
where $X$ denotes the Pauli-$X$ (bit-flip) operator. Consequently, the completely positive trace-preserving (CPTP) map induced by the symmetric Unruh transformation takes the compact Kraus form
\begin{equation}
\begin{aligned}
\mathcal{E}_{\mathrm{sym}}(\rho)
&=
K_0(r)\rho K_0^\dagger(r)
+
K_1(r)\rho K_1^\dagger(r)\\
&=
[1-p(r)]\rho + p(r)X\rho X.
\end{aligned}
\end{equation}


\section{Entanglement: revisited versus standard Unruh effect}
\label{EUEARUE}

In this section, we investigate the entanglement properties of the revisited Unruh effect and compare them with those of the standard Unruh effect. To investigate the entanglement properties of the resulting state, we employ the negativity and logarithmic negativity measures. For this purpose, we first calculate the partial transpose of the reduced density matrix with respect to subsystem $A$:

\begin{equation}
\rho_{AR}^{T_A}
=
\frac{1}{2}
\begin{pmatrix}
\cos^2 r & 0 & 0 & \sin^2 r\\[4pt]
0 & \sin^2 r & \cos^2 r & 0\\[4pt]
0 & \cos^2 r & \sin^2 r & 0\\[4pt]
\sin^2 r & 0 & 0 & \cos^2 r
\end{pmatrix}.
\end{equation}

The eigenvalues of the partially transposed density matrix are obtained as

\begin{equation}
\lambda_1
=
\lambda_2
=
\frac{1}{2},
\qquad
\lambda_3
=
\frac{1}{2}\cos 2r,
\qquad
\lambda_4
=
-\frac{1}{2}\cos 2r.
\end{equation}

The negativity is defined in terms of the trace norm of the partial transpose as\cite{vidal2002computable}:

\begin{equation}
N(\rho)
=
\frac{\left\|\rho^{T_A}\right\|_1-1}{2},
\label{NegF2}
\end{equation}
where $\left\|\rho^{T_A}\right\|_1$ denotes the trace norm of the partially transposed density matrix. For a two-qubit system, the negativity ranges from $0$ to $1/2$. For the bipartition $AR$, the negativity is obtained as
\begin{equation}
\tilde{N}(r)
=
\frac{1}{2}\cos(2r).
\end{equation}

\begin{figure}[H]
    \centering

    \begin{subfigure}{0.45\textwidth}
        \centering
        \includegraphics[width=\linewidth]{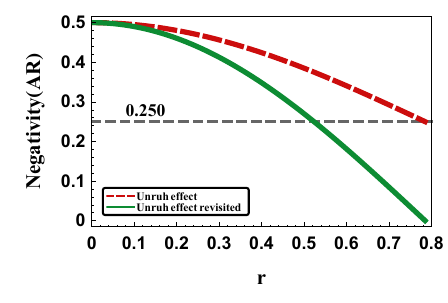}
        \caption*{(a)}
        \label{fig:sym_neg}
    \end{subfigure}
    \hfill
    \begin{subfigure}{0.45\textwidth}
        \centering
        \includegraphics[width=\linewidth]{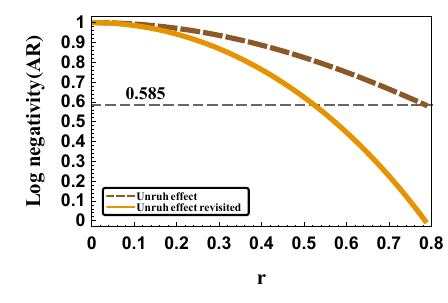}
        \caption*{(b)}
        \label{fig:sym_logneg}
    \end{subfigure}

    \caption{(a) Negativity and (b) logarithmic negativity of the pair $(A,R)$ versus the acceleration parameter $r$ for the Unruh effect revisited (solid lines) and the standard Unruh effect (dashed lines).}
\label{fig:symmetric_unruh_entanglement}
\end{figure}

Please note that throughout the paper, all tilded relations correspond to the revisited Unruh effect.
For the standard Unruh effect, the negativity has also been investigated previously and is obtained as\cite{mehri2011pseudo}:
\begin{equation}
N(r)
=
\frac{1}{2}\cos^{2}(r).
\end{equation}
The logarithmic negativity is defined in terms of the negativity as\cite{vidal2002computable}:
\begin{equation}
E_N(\rho)
=
\log_2\left(1+2N(\rho)\right),
\label{LogNeg}
\end{equation}
which ranges from $0$ to $1$. For the present state, the corresponding logarithmic negativity is
\begin{equation}
\tilde{E}_N(r)
=
\log_2\left(1+\cos(2r)\right).
\end{equation}
Similarly, for the standard Unruh effect, the logarithmic negativity has been studied previously and is given by\cite{mehri2011pseudo}:
\begin{equation}
E_N(r)
=
\log_2\left(1+\cos^{2}(r)\right).
\end{equation}
The behavior of the negativity and logarithmic negativity as functions of the acceleration parameter $r$ is illustrated in Figure~\ref{fig:symmetric_unruh_entanglement}. For a better comparison with the standard Unruh effect, the corresponding results, $N_{\mathrm{U}}(r)$ and $E_{N,\mathrm{U}}(r)$, are also included in the figure. In the inertial limit, $r=0$, the system is maximally entangled and the corresponding values are $N=0.5$ and $E_N=1$. As the acceleration increases, the entanglement decreases continuously and eventually vanishes in the infinite-acceleration limit, $r=\pi/4$, where $N=0$ and $E_N=0$.
The vanishing values of both $N$ and $E_N$ indicate that the resulting state becomes separable in the infinite-acceleration limit.
This behavior highlights the fundamental difference between the symmetric and asymmetric Unruh transformations. In the symmetric case, the entanglement is completely destroyed at infinite acceleration due to the equal contribution of excitation and de-excitation processes. However, in the standard asymmetric Unruh effect, a finite amount of entanglement survives even in the infinite acceleration limit.

The entanglement properties of the remaining bipartitions can be analyzed following the same procedure. Starting from the tripartite wave function in Eq.~(\ref{rhoarsy}), tracing out subsystem $R$ and subsystem $A$, respectively, yields the reduced density matrices $\rho_{A,\overline{R}}^{(\mathrm{sym})}$ and $\rho_{R,\overline{R}}^{(\mathrm{sym})}$. These two reduced density matrices exhibit identical entanglement properties, such that the bipartitions $A\overline{R}$ and $R\overline{R}$ have the same negativity and logarithmic negativity. Applying the partial-transpose criterion and using the definitions of negativity and logarithmic negativity given in Eqs.~(\ref{NegF2}) and~(\ref{LogNeg}), respectively, both bipartitions are found to have vanishing entanglement throughout the entire allowed range $0\leq r\leq\pi/4$. In particular, the negativities are given by
\begin{equation}
\tilde{N}_{A\overline{R}}(r)=\tilde{N}_{R\overline{R}}(r)=0,
\end{equation}
and the corresponding logarithmic negativities are
\begin{equation}
\tilde{E}_{N,A\overline{R}}(r)=\tilde{E}_{N,R\overline{R}}(r)=0.
\end{equation}
Thus, within the symmetric transformation considered here, no bipartite entanglement is generated between Alice and the inaccessible Rindler sector, nor between the two Rindler sectors, for any value of the acceleration parameter $r$.

In contrast, for the standard Unruh effect, the bipartitions
$A\overline{R}$ and $R\overline{R}$ are entangled. Their negativities
and logarithmic negativities have also been investigated previously\cite{mehri2011pseudo}.
For the bipartition $(A,\overline{R})$, the negativity is given by
\begin{equation}
N_{A\overline{R}}(r)
=
\frac{1}{2}\sin^{2}(r).
\end{equation}
The corresponding logarithmic negativity is
\begin{equation}
E_{N,A\overline{R}}(r)
=
\log_2\left(1+\sin^{2}(r)\right).
\end{equation}

Similarly, for the bipartition $(R,\overline{R})$, the negativity
is given by
\begin{equation}
N_{R\overline{R}}(r)
=
\frac{\sqrt{1+\sin^{2}(2r)}-1}{4}.
\end{equation}
The corresponding logarithmic negativity reads
\begin{equation}
E_{N,R\overline{R}}(r)
=
\log_2\left(
\frac{1+\sqrt{1+\sin^{2}(2r)}}{2}
\right).
\end{equation}

\begin{figure*}[t]
    \centering

    \begin{subfigure}{0.40\textwidth}
        \centering
        \includegraphics[width=\linewidth]{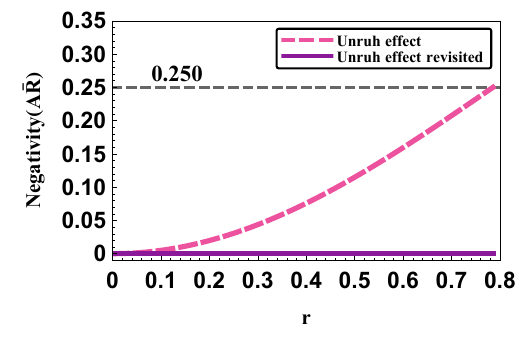}
        \caption*{(a)}
        \label{fig:sym_neg_ARbar}
    \end{subfigure}
    \hspace{0.03\textwidth}
    \begin{subfigure}{0.40\textwidth}
        \centering
        \includegraphics[width=\linewidth]{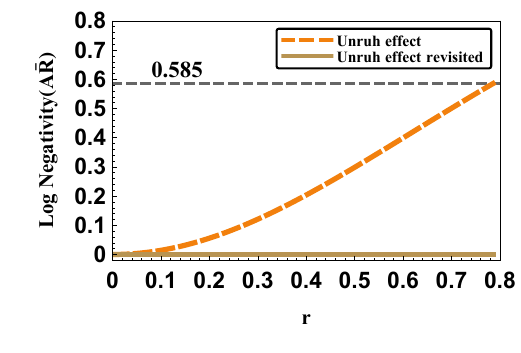}
        \caption*{(b)}
        \label{fig:sym_logneg_ARbar}
    \end{subfigure}

    \vspace{0.5cm}

    \begin{subfigure}{0.40\textwidth}
        \centering
        \includegraphics[width=\linewidth]{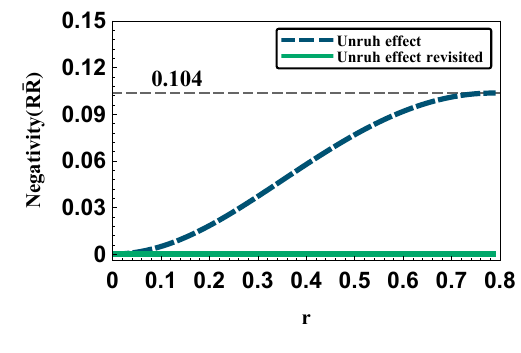}
        \caption*{(c)}
        \label{fig:sym_neg_RRbar}
    \end{subfigure}
    \hspace{0.03\textwidth}
    \begin{subfigure}{0.40\textwidth}
        \centering
        \includegraphics[width=\linewidth]{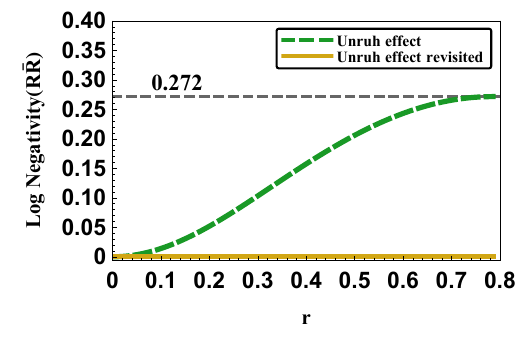}
        \caption*{(d)}
        \label{fig:sym_logneg_RRbar}
    \end{subfigure}

    \caption{Negativity and logarithmic negativity versus $r$ for the Unruh effect revisited (solid lines) and the standard Unruh effect (dashed lines): (a),(b) pair $A\overline{R}$; (c),(d) pair $R\overline{R}$.}
    \label{fig:sym_ent_ARbar_RRbar}
\end{figure*}

All four quantities vanish in the inertial limit, $r=0$, and increase monotonically with the acceleration parameter. In the infinite-acceleration limit, $r=\pi/4$, they reach the finite values. The behavior of the negativity and logarithmic negativity of the bipartitions $A\overline{R}$ and $R\overline{R}$ as functions of the acceleration parameter $r$ is shown in Fig.~\ref{fig:sym_ent_ARbar_RRbar}. For a better comparison with the standard Unruh effect, its results are also included in the figure.

\section{Quantum mutual information: revisited versus standard}
\label{QMIUEARUE}
Having characterized the entanglement behavior of the symmetric state, we now turn to the quantum mutual information to examine the total correlations between subsystems. Unlike entanglement measures, which quantify only the nonclassical part of correlations, quantum mutual information captures the total amount of correlations, including both quantum and classical contributions. Therefore, it provides a complementary characterization of how the correlations between Alice and the accelerated observer are affected by the symmetric Unruh transformation.
The quantum mutual information between two subsystems $A$ and $R$ is defined as:
\begin{equation}
I(X:Y)
=
S(\rho_X)
+
S(\rho_Y)
-
S(\rho_{XY}),
\label{MI_sym_definition}
\end{equation}
where $S(\rho)$ denotes the von Neumann entropy of a density matrix $\rho$, defined by
\begin{equation}
S(\rho)
=
-\operatorname{Tr}\left(\rho\log_2\rho\right)
=
-\sum_i \lambda_i\log_2\lambda_i,
\label{entropy_sym}
\end{equation}
where $\lambda_i$ are the eigenvalues of $\rho$. It is clear that the quantum mutual information is symmetric, i.e.,
\begin{equation}
I(X:Y)=I(Y:X).
\end{equation}
For a bipartite state $\rho_{AR}$, the reduced density matrices of the individual subsystems are obtained by taking the partial trace over the complementary subsystem,
\begin{equation}
\rho_A
=
\operatorname{Tr}_{R}\left(\rho_{AR}\right),
\qquad
\rho_R
=
\operatorname{Tr}_{A}\left(\rho_{AR}\right).
\label{reduced_states_MI_sym}
\end{equation}
Accordingly, $S(\rho_A)$ represents the von Neumann entropy of subsystem $A$, obtained after tracing out subsystem $R$, while $S(\rho_R)$ represents the von Neumann entropy of subsystem $R$, obtained after tracing out subsystem $A$. The term $S(\rho_{AR})$ is the von Neumann entropy of the complete bipartite state.

Substituting the Eqs.\eqref{rhoarsy}, \eqref{MI_sym_definition} and \eqref{entropy_sym} we obtain:
\begin{equation}
\tilde{I}(A:R)
=
2
+
\sin^2 r\log_2\left(\sin^2 r\right)
+
\cos^2 r\log_2\left(\cos^2 r\right).
\label{MI_sym_final}
\end{equation}

Here, $I(A:R)$ quantifies the total correlations shared between subsystems $A$ and $R$, where $A$ denotes Alice's subsystem and $R$ denotes the accessible Rindler region associated with the accelerated observer. The quantum mutual information can be expressed in the compact form
\begin{equation}
\tilde{I}(A:R)
=
2-h\left(\sin^2 r\right)
.
\label{MI_sym_compact}
\end{equation}
Here, $h(x)$ denotes the binary entropy function, which is defined as
\begin{equation}
h(x)
=
-x\log_2 x-(1-x)\log_2(1-x).
\label{binary_entropy}
\end{equation}
and relation $h(x)=h(1-x)$ has been used to obtain this expression.
For the standard Unruh effect, the quantum mutual information between $A$ and $R$ has also been investigated previously and is obtained as\cite{mehri2011pseudo}:
\begin{equation}
I(A:R)
=
1
+
h\left(\frac{\cos^{2} r}{2}\right)
-
h\left(\frac{\sin^{2} r}{2}\right).
\label{MI_standard_AR}
\end{equation}
Using Eq.~\eqref{MI_sym_final}, the quantum mutual information as a function of the acceleration parameter $r$ is shown in panel (a) of Fig.~\ref{fig:sym_mutual_information}. For a better comparison with the standard Unruh effect, the corresponding quantum mutual information is also included in the figure. It can be seen that, in the zero-acceleration limit, corresponding to $r=0$, the quantum mutual information attains its maximum value of 2. As the acceleration increases, it decreases monotonically, reaching a value of 1 in the infinite-acceleration limit, corresponding to $r=\pi/4$, even though the entanglement has completely vanished at this point. The standard Unruh effect exhibits the same limiting values, with the quantum mutual information decreasing from 2 at zero acceleration to 1 at infinite acceleration. However, the two cases differ in their behavior at intermediate accelerations: the quantum mutual information decreases more rapidly in the revised Unruh framework than in the standard Unruh effect. This difference highlights the distinct ways in which acceleration affects the correlations between Alice and Rob in the two frameworks.

Following the same procedure as above, we now use Eqs.~\eqref{MI_sym_definition},\eqref{entropy_sym} to evaluate the total correlations for the remaining two pairs of subsystems, $(A,\overline{R})$ and $(R,\overline{R})$, obtained by tracing out $R$ or $A$, Because, as established previously, the two reduced density matrices coincide, 
\begin{equation} 
\rho_{A,\overline{R}}^{(\mathrm{sym})}=\rho_{R,\overline{R}}^{(\mathrm{sym})}, 
\label{rhoARbarsy} 
\end{equation} 
their von Neumann entropies, and hence their quantum mutual informations, are trivially equal to one another as well: 
\begin{equation}\label{MI_ARbar}
\begin{aligned}
\tilde{I}(A:\overline{R})=\tilde{I}(R:\overline{R})
&=
-\sin^{2} r\log_2\left(\sin^{2} r\right)\\
&\quad
-\cos^{2} r\log_2\left(\cos^{2} r\right),
\end{aligned}
\end{equation}
Using the binary entropy function introduced in Eq.~\eqref{binary_entropy}, this result can be written in the compact form 
\begin{equation} 
\tilde{I}(A:\overline{R})=\tilde{I}(R:\overline{R})=h\left(\sin^{2} r\right). 
\label{MI_ARbar_compact} 
\end{equation} 
For the standard Unruh effect, the mutual informations of the remaining two pairs have also been investigated previously and are given by\cite{mehri2011pseudo}:
\begin{equation}
\begin{aligned}
I(A:\overline{R})
&=
1
+
h\left(\frac{\sin^{2} r}{2}\right)
-
h\left(\frac{\cos^{2} r}{2}\right),\\
I(R:\overline{R})
&=
h\left(\frac{\cos^{2} r}{2}\right)
+
h\left(\frac{\sin^{2} r}{2}\right)
-
1.
\end{aligned}
\label{MI_standard_ARbar_RRbar}
\end{equation}
\begin{widetext}

Comparing Eqs.~\eqref{MI_ARbar_compact} and
\eqref{MI_sym_compact}, we observe that the sum of the quantum
mutual information between Alice and Rob and that between Alice
and the inaccessible mode remains constant at 2 for all values
of $r$. Thus, the total correlations are redistributed between
the accessible pair $(A,R)$ and the pair involving the
inaccessible mode $\overline{R}$, while their sum remains
constant.

At $r=0$, the quantum mutual information between Alice and the
inaccessible mode vanishes. This is consistent with the fact
that the reduced density matrix of Alice and the inaccessible
mode becomes a product state in the inertial limit. As the
acceleration increases, their quantum mutual information grows
monotonically, reaching its maximum value of 1 in the
infinite-acceleration limit. This behavior indicates that, by
the time the accessible entanglement has been completely
destroyed, half of the correlations initially present between
Alice and Rob have been transferred to the pair
$(A,\overline{R})$. Equivalently, the same conclusion holds for
the pair $(R,\overline{R})$. Consequently, in the
infinite-acceleration limit, the quantum mutual information
equals 1 for both pairs.

The same sum rule also holds for the standard Unruh effect,
and the quantum mutual information between Alice and Rob and
that between Alice and the inaccessible mode both reach 1 in
the infinite-acceleration limit. However, the quantum mutual
information between Alice and the inaccessible mode increases
more slowly in the standard Unruh effect than in the revised
framework. Moreover, the quantum mutual information between
Rob and the inaccessible mode in the standard Unruh effect
saturates at approximately $0.623$ in the infinite-acceleration
limit, whereas the corresponding quantity in the revised
Unruh effect reaches 1.

The behavior of the quantum mutual information between Rob and
the inaccessible mode, as well as that between Alice and the
inaccessible mode, as functions of the acceleration parameter
$r$, is depicted in panels (b) and (c), respectively, of
Fig.~\ref{fig:sym_mutual_information}. The corresponding
results for the standard Unruh effect are also included for
comparison.

\begin{figure}[H]
    \centering

    \begin{subfigure}{0.32\textwidth}
        \centering
        \includegraphics[width=\linewidth]{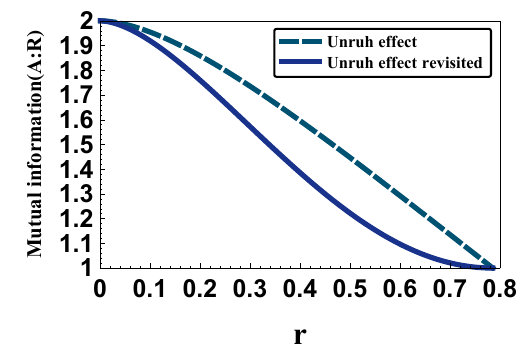}
        \caption*{(a)}
        \label{fig:sym_mi_AR}
    \end{subfigure}
    \hfill
    \begin{subfigure}{0.32\textwidth}
        \centering
        \includegraphics[width=\linewidth]{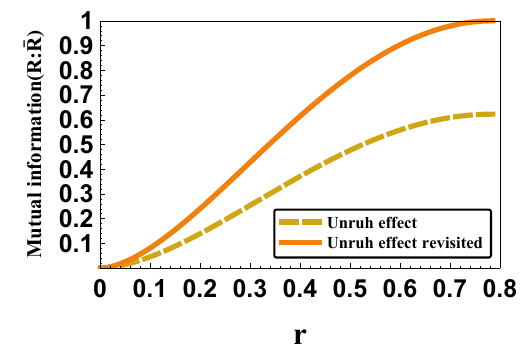}
        \caption*{(b)}
        \label{fig:sym_mi_RRbar}
    \end{subfigure}
    \hfill
    \begin{subfigure}{0.32\textwidth}
        \centering
        \includegraphics[width=\linewidth]{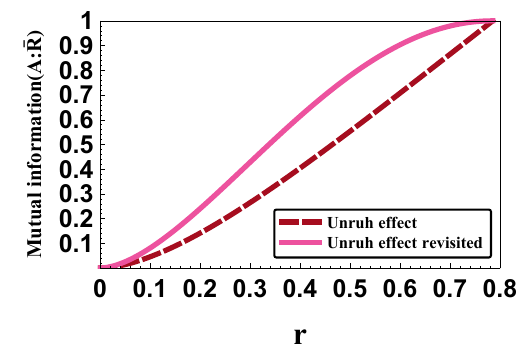}
        \caption*{(c)}
        \label{fig:sym_mi_ARbar}
    \end{subfigure}
    \caption{Quantum mutual information as a function of the acceleration parameter $r$ for the Unruh effect revisited (solid lines), compared with the standard Unruh effect (dashed lines): (a) mutual information $(A\!:\!R)$; (b) mutual information $(R\!:\!\bar{R})$; (c) mutual information $(A\!:\!\bar{R})$.}
    \label{fig:sym_mutual_information}
\end{figure}
\end{widetext}


\section{Quantum discord: revisited versus standard}
\label{QDUEARUE}
Having examined the total correlations through the quantum mutual information, we now turn to quantum discord, which quantifies the genuinely quantum part of the correlations that cannot be accounted for by classical correlations alone. Quantum discord was introduced by Ollivier and Zurek and is based on the difference between two classically equivalent descriptions of mutual information, which become inequivalent in the quantum regime. In particular, quantum discord measures the quantum correlations that may persist even in the absence of entanglement.

For a bipartite state $\rho_{AR}$, the quantum discord when the measurement is performed on subsystem $R$ is defined as
\begin{equation}
D(A:R)=I(A:R)-C(A:R),
\label{D_def}
\end{equation}
where $I(A:R)$ is the quantum mutual information defined in Eq.~\eqref{MI_sym_definition}, and $C(A:R)$ denotes the maximal classical correlation obtainable by performing a local measurement on $R$. More explicitly,
\begin{equation}
C(A:R)
=
\max_{\{\Pi_k^R\}}
\left[
S(\rho_A)
-
\sum_k p_k S(\rho_{A|k})
\right],
\label{C_def}
\end{equation}
where $\{\Pi_k^R\}$ represents a complete set of projective measurements on $R$, $p_k$ is the probability of obtaining the outcome $k$, and $\rho_{A|k}$ is the corresponding conditional state of subsystem $A$. The discord in the opposite direction, $D(R:A)$, is obtained by performing the measurement on $A$. In general, quantum discord is asymmetric, so that $D(A:R)\neq D(R:A)$.

We first consider the pair $(A,R)$ described by the reduced density matrix $\rho_{AR}^{(\mathrm{sym})}$ given in Eq.~\eqref{rhoarsy}. As established from the corresponding mutual information in Eq.~\eqref{MI_sym_final}, and by optimizing the local measurement over the measurement basis, the optimal measurement is found to be along the $\sigma_x$ direction. This measurement projects the other subsystem onto pure conditional states for every value of $r$. Consequently, the classical correlation reaches its maximum value,
\begin{equation}
\tilde{C}(A:R)=\tilde{C}(R:A)=1.
\label{Csym}
\end{equation}
Combining this result with the quantum mutual information of Eq.~\eqref{MI_sym_final}, the quantum discord is obtained as
\begin{equation}
\begin{aligned}
\tilde{D}(A:R)=\tilde{D}(R:A)
&=
1+\cos^2 r\log_2\left(\cos^2 r\right)\\
&\quad
+\sin^2 r\log_2\left(\sin^2 r\right).
\end{aligned}
\label{Dsym_final}
\end{equation}
Using the binary entropy function , this result can be written in the compact form
\begin{equation}
\tilde{D}(A:R)=\tilde{D}(R:A)
=
1-h\left(\sin^2 r\right).
\label{Dsym_compact}
\end{equation}
The equality of the two discord directions follows from the $A\leftrightarrow R$ exchange symmetry of the reduced state.

We now consider the two remaining pairs, $(A,\overline{R})$ and $(R,\overline{R})$, obtained from the tripartite state $|\Phi^{+}\rangle^{(\mathrm{sym})}_{A,R,\overline{R}}$ by tracing out $R$ and $A$, respectively. Owing to the $A\leftrightarrow R$ symmetry of the tripartite state, the two reduced states $\rho^{(\mathrm{sym})}_{A,\overline{R}}$ and $\rho^{(\mathrm{sym})}_{R,\overline{R}}$ are identical, so it suffices to evaluate the discord for one of them.

To obtain the quantum discord, we maximize the classical correlation over all local projective measurements. For a measurement on $\overline{R}$, the optimal direction is along the $\sigma_x$ axis. The resulting conditional states of $A$ (or $R$) have Bloch-vector length $\sin 2r$, which gives
\begin{equation}
\tilde{C}(A:\overline{R}) = \tilde{C}(R:\overline{R}) = 1 - h\!\left(\frac{1+\sin 2r}{2}\right).
\label{C_ARbar}
\end{equation}
Substituting Eqs.~\eqref{MI_ARbar} and \eqref{C_ARbar} into the definition of quantum discord, Eq.~\eqref{D_def}, we obtain
\begin{equation}
\tilde{D}(A:\overline{R}) = \tilde{D}(R:\overline{R}) = h(\sin^2 r) + h\!\left(\frac{1+\sin 2r}{2}\right) - 1.
\label{D_ARbar}
\end{equation}

In the opposite direction, when the measurement is performed on $A$ (or $R$), the conditional state of $\overline{R}$ is pure for every value of $r$. The classical correlation then equals the mutual information, $C(\overline{R}:A) = C(\overline{R}:R) = h(\sin^2 r)$, and therefore
\begin{equation}
\tilde{D}(\overline{R}:A) = \tilde{D}(\overline{R}:R) = 0.
\label{D_barR_zero}
\end{equation}

Thus, among the six possible quantum discords associated with the subsystems $A$, $R$, and $\overline{R}$, two vanish identically, while the remaining four, $\tilde{D}(A:R)$, $\tilde{D}(R:A)$, $\tilde{D}(A:\overline{R})$, and $\tilde{D}(R:\overline{R})$, are generally nonzero.

\begin{figure}[t]
    \centering
    \begin{subfigure}[b]{0.45\textwidth}
        \centering
        \includegraphics[width=\textwidth]{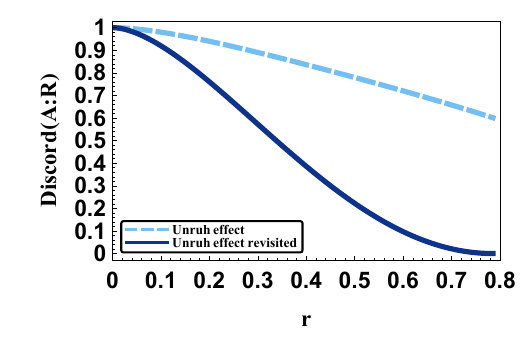}
        \caption{}
        \label{fig:cmp_AR_a}
    \end{subfigure}

    \vspace{-2mm}

    \begin{subfigure}[b]{0.45\textwidth}
        \centering
        \includegraphics[width=\textwidth]{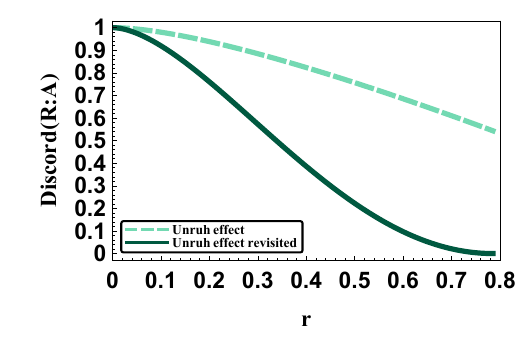}
        \caption{}
        \label{fig:cmp_AR_b}
    \end{subfigure}
    \caption{Quantum discords of the pair $(A,R)$ for the standard Unruh effect (dashed lines) and the revisited Unruh effect (solid lines) versus the acceleration parameter $r$: (a) Discord $(A:R)$ and (b) Discord $(R:A)$.}
    \label{fig:cmp_AR}
\end{figure}

For comparison, we recall the quantum discords of the standard Unruh effect, which have been studied previously\cite{mehri2011pseudo}. Using the same convention, with $R$ and $\overline{R}$ denoting the accessible and inaccessible Rindler modes, $h(x)$ the binary entropy, and
\begin{equation}
\kappa(r)=\sqrt{1-\frac{\sin^2 2r}{4}},
\label{kappa_def}
\end{equation}
the discords of the pair $(A,R)$ are
\begin{align}
D(A:R) &= h\!\left(\frac{\cos^2 r}{2}\right)
         - h\!\left(\frac{\sin^2 r}{2}\right)
         + h\!\left(\frac{1+\cos r}{2}\right),
\label{D_AR_std}\\[4pt]
D(R:A) &= 1 - h\!\left(\frac{\sin^2 r}{2}\right)
         + h\!\left(\frac{1+\kappa(r)}{2}\right).
\label{D_RA_std}
\end{align}
Unlike the revisited effect, the two directions differ in the standard case, $D(A:R)\neq D(R:A)$. The comparison between the two effects for this pair is shown in Fig.~\ref{fig:cmp_AR}.

\begin{figure*}[t]
    \centering
    \begin{subfigure}[b]{0.40\textwidth}
        \centering
        \includegraphics[width=\linewidth]{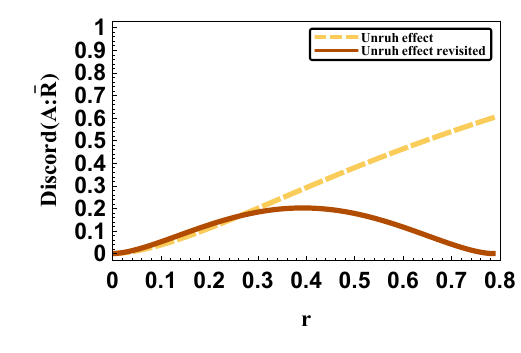}
        \caption{}
        \label{fig:cmp_ARb_a}
    \end{subfigure}%
    \hspace{1mm}%
    \begin{subfigure}[b]{0.40\textwidth}
        \centering
        \includegraphics[width=\linewidth]{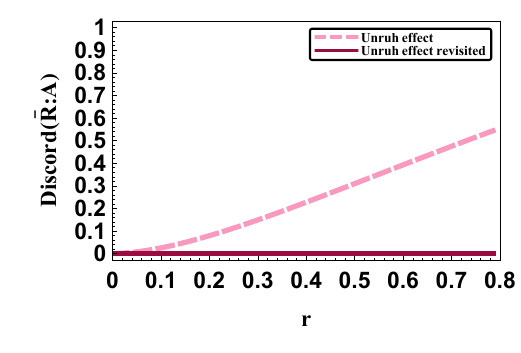}
        \caption{}
        \label{fig:cmp_ARb_b}
    \end{subfigure}

    \vspace{-1mm}

    \begin{subfigure}[b]{0.40\textwidth}
        \centering
        \includegraphics[width=\linewidth]{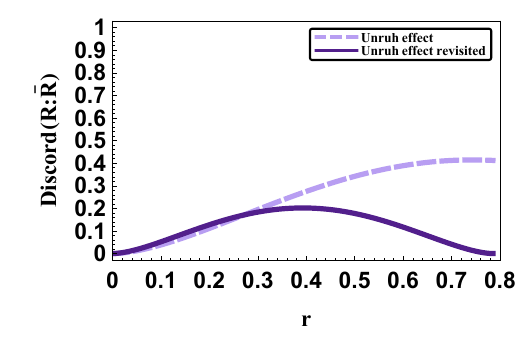}
        \caption{}
        \label{fig:cmp_ARb_c}
    \end{subfigure}%
    \hspace{1mm}%
    \begin{subfigure}[b]{0.40\textwidth}
        \centering
        \includegraphics[width=\linewidth]{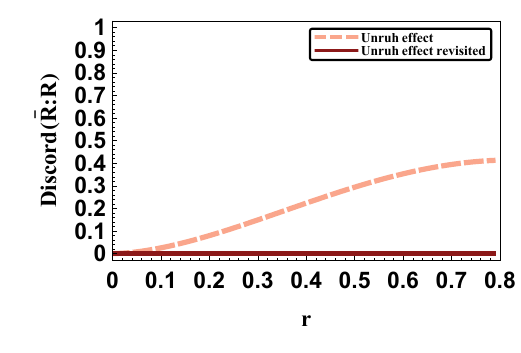}
        \caption{}
        \label{fig:cmp_ARb_d}
    \end{subfigure}
    \caption{Quantum discords of the pairs $(A,\overline{R})$ and $(R,\overline{R})$ for the standard Unruh effect (dashed lines) and the revisited Unruh effect (solid lines) versus the acceleration parameter $r$: (a) Discord $(A:\overline{R})$, (b) Discord $(\overline{R}:A)$, (c) Discord $(R:\overline{R})$, and (d) Discord $(\overline{R}:R)$. }
    \label{fig:cmp_ARb}
\end{figure*}

For the pair $(A,\overline{R})$, the standard Unruh discords are
\begin{align}
D(A:\overline{R}) &= h\!\left(\frac{\sin^2 r}{2}\right)
         - h\!\left(\frac{\cos^2 r}{2}\right)
         + h\!\left(\frac{1+\sin r}{2}\right),
\label{D_ARb_std}\\[4pt]
D(\overline{R}:A) &= 1 - h\!\left(\frac{\cos^2 r}{2}\right)
         + h\!\left(\frac{1+\kappa(r)}{2}\right),
\label{D_RbA_std}
\end{align}
and for the pair $(R,\overline{R})$ they are
\begin{align}
D(R:\overline{R}) &= h\!\left(\frac{\sin^2 r}{2}\right) - 1
         + h\!\left(\frac{1+\sin r}{2}\right),
\label{D_RRb_std}\\[4pt]
D(\overline{R}:R) &= h\!\left(\frac{\cos^2 r}{2}\right) - 1
         + h\!\left(\frac{1+\cos r}{2}\right).
\label{D_RbR_std}
\end{align}
In contrast to the revisited effect, where $D(\overline{R}:A)$ and $D(\overline{R}:R)$ vanish identically, all four discords above are generally nonzero. The comparison for these two pairs is displayed in Fig.~\ref{fig:cmp_ARb}.

Overall, the revisited transformation changes not only the magnitude of the quantum discords but also their structure: two of the six discords vanish identically, and the discords of the pair $(A,R)$ become direction-independent.

In the standard Unruh effect, the quantum discord between Alice and the accessible Rindler mode can remain nonzero in the infinite-acceleration limit, even after the entanglement has vanished. In contrast, for the symmetric transformation considered here, the entanglement and the quantum discord of the pair $(A,R)$ vanish simultaneously at $r=\pi/4$. Hence, the revisited transformation leaves no residual quantum correlation between Alice and the accessible Rindler mode at infinite acceleration.


\section{Conclusion}\label{Conclusion}

We have revisited the Unruh effect from the standpoint of relativistic quantum
information, with the aim of understanding why, under the standard asymmetric
transformation for a Dirac field, entanglement and quantum discord between an
inertial and a uniformly accelerated observer do not vanish at infinite
acceleration. We proposed a symmetric Unruh-like map in which both the
Minkowski vacuum and the one-particle state exchange amplitude with the
inaccessible Rindler region. For an initial Bell state, the reduced state of
the accessible pair is generated by a bit-flip channel whose flip probability
is the Fermi-Dirac occupation at the Unruh temperature. This should be
contrasted with the standard transformation, which acts as a one-sided
amplitude-damping channel. In the standard case only the vacuum component is
excited, whereas in the symmetric case excitation and de-excitation enter on
the same footing, as expected for a thermal environment.
 
For the accessible pair, the negativity and the logarithmic negativity decrease
monotonically from their maximal values in the inertial limit and vanish only at
infinite acceleration. In the standard Unruh effect, the same limit leaves a finite negativity and a finite logarithmic negativity. The quantum discord of the accessible pair vanishes at the same
point, and it is identical in both measurement directions, so that entanglement
and discord disappear simultaneously. The mutual information of this pair
decreases from two bits to one bit and not to zero, which identifies the
surviving correlation as classical. The standard Unruh effect reaches the same
final value of one bit, but with a different dependence on the acceleration at
intermediate values, and it retains quantum correlations in addition.
 
The inaccessible sector shows a complementary behavior. The sum of the mutual
informations of the accessible pair and of Alice with the inaccessible mode is
constant for all accelerations, so the total correlations are redistributed
between the accessible and inaccessible sectors and not lost.  Despite carrying this amount of total correlation, the
pairs involving the inaccessible mode are separable for every acceleration,
while in the standard case they are entangled. Their correlations are therefore
of a different nature: of the six pairwise discords, the two with the
inaccessible mode as the measured party vanish identically, because measuring
either accessible qubit leaves the inaccessible mode in a pure conditional
state, and the remaining four are generally nonzero.
 
These results suggest that the survival of quantum correlations at infinite
acceleration is not forced by fermionic statistics or by the horizon alone, but
depends on how the Minkowski-Rindler map distributes amplitude between the
accessible and inaccessible sectors. A balanced, two-way exchange, which is the
structure expected of a thermalization process, removes all quantum
correlations of the accessible pair, whereas the asymmetric standard map does
not. The channel language used here makes this comparison operational and
provides a common framework for analyzing different constructions of the
Unruh effect.
 
Several limitations should be kept in mind. The symmetric map is a model, and
it is not derived from the Bogoliubov transformation of the Dirac field.
Moreover, it is expressed within a single-mode description, and it is known
that the standard predictions depend on the choice of Unruh modes and on the
single-mode approximation. Whether the symmetric map corresponds to a
particular choice of Unruh modes, or to a physically distinct regime, remains an open question. 
\bibliography{ref}

\end{document}